\documentclass[]{sig-alternate-per}
\pdfoutput=1
\usepackage{balance}

\usepackage{hyperref}
\hypersetup{
	colorlinks = true,
linkcolor=blue,
citecolor=blue,
urlcolor=blue,
pdftitle=SW-QPS}
\usepackage[lined,boxed,linesnumbered,vlined]{algorithm2e}

\usepackage{etoolbox}
\usepackage{comment}
\def\equationautorefname~#1\null{%
  (#1)\null
}

\SetKwInput{Problem}{problem}
\SetKwInput{Input}{input}
\SetKwInput{Output}{output}
\SetAlgoSkip{smallskip} 
\SetAlgoInsideSkip{smallskip} %
\SetKwComment{tcp}{\#\ }{}
\SetKwComment{tcc}{\#\ }{}
\newcommand{\ie}{{\em i.e.,} }
\ifcsdef{sitemize}{}{

}

\ifcsdef{senumerate}{}{

}

\ifcsdef{spenumerate}{}{

}

\ifcsdef{slenumerate}{}{

}

\usepackage{cite}
\usepackage[group-separator={,},group-minimum-digits=4]{siunitx}

\usepackage{dblfloatfix}
\def\fourfull{0.958\textwidth}
\begin{document}
\title{Sliding-Window QPS (SW-QPS):  A Perfect Parallel Iterative Switching Algorithm for Input-Queued Switches}


\def\saddr{Georgia Tech}
\numberofauthors{3}
\author{
\alignauthor Jingfan Meng\\
 \affaddr{\saddr}\\
 \email{jmeng40@gatech.edu}
\alignauthor Long Gong\\
 \affaddr{\saddr}\\
 \email{gonglong@gatech.edu}
 \alignauthor Jun (Jim) Xu\\
 \affaddr{\saddr}\\
 \email{jx@cc.gatech.edu}
}

\maketitle
\begin{abstract}

In this work, we first propose a parallel batch switching algorithm called Small-Batch Queue-Proportional Sampling (SB-QPS).  
Compared to other batch switching algorithms, SB-QPS significantly reduces the batch size without sacrificing the throughput performance and hence has much lower delay 
when traffic load is light to moderate.  It also 
achieves
the lowest possible time complexity of $O(1)$ per matching computation per port, via parallelization. 
We then propose another algorithm called Sliding-Window QPS (SW-QPS).  SW-QPS retains and enhances 
all benefits of SB-QPS, and reduces the batching delay to zero via 
a novel switching framework called sliding-window switching. 
In addition, SW-QPS computes matchings of much higher qualities, 
as measured by the resulting throughput and delay performances, than QPS-1, the state-of-the-art regular switching algorithm 
that builds upon the same 
underlying bipartite matching algorithm. 

%
\end{abstract}


\section{Introduction}\label{sec:intro}

Many present day switching systems in Internet routers and data-center switches employ 
an input-queued crossbar to interconnect their input ports and output ports. 
In an $N\times N$ input-queued crossbar switch, 
each input port has $N$ Virtual Output Queues
(VOQs). A VOQ $j$ at input port $i$ serves as a buffer for the packets
going into input port $i$ destined for output port $j$. The use of VOQs solves the Head-of-Line
(HOL) blocking issue~\cite{KarolHluchyjMorgan1987HOL}, which
severely limits the throughput of input-queued switches.

In an $N\times N$ input-queued crossbar switch, each input port can be connected to only one output
port and vice versa in each switching cycle or time slot.
Hence, in every time slot, the switch needs to compute
a one-to-one \textit{matching} ({\it i.e.}, the crossbar schedule) between input and output ports .
A major research challenge
of designing high-link-rate 
switches with a large number of ports (called {\it high-radix}~\cite{Cakir2016HighRadixCrossbar}) 
is to develop switching algorithms that can
compute ``high quality'' matchings -- those that result in
high switch throughput and low queueing
delays for packets -- in a short time slot.

\subsection{Existing Approaches}\label{subsec:existing}

While 
many switching algorithms have been proposed for input-queued switches, 
they either have a (relatively) high time complexity that 
prevents a matching computation from being completed in a short time slot, or cannot produce high-quality matchings 
that translate into 
excellent throughput and delay performances.   For example, the widely-used iSLIP algorithm~\cite{McKeown99iSLIP} 
can empirically achieve over 80\% throughputs under most of traffic patterns, as will be shown in~\autoref{subsec:throughput-results}.   
However, even with 
a parallel iterative implementation, its time complexity per port is $O(\log^2 N)$, which is still
too high when the switch size $N$ is large and the time slot is short (say a few nanoseconds long).






It is possible to improve the quality of the matching without increasing the time complexity of the switching algorithm using a strategy 
called batching~\cite{Aggarwal2003EdgeColoring,Neely2007FrameBased,Wang2018ParallelEdgeColoring}. 
Unlike in a regular switching algorithm, where a matching decision is computed for every time slot, 
in a batch switching algorithm, multiple (say $T$) consecutive time slots are grouped as
a batch and these $T$ matching decisions are batch-computed.  Hence, in a batch switching algorithm, each of the $T$ matchings-under-compuation in a batch
has a period of $T$ time slots to find opportunities to have the quality of the matching 
improved by the underlying bipartite matching algorithm, 
whereas in a regular switching algorithm, each matching has only a single time slot to find such opportunities.  
As a result, a batch switching algorithm can usually produce matchings of 
higher qualities than a regular switching algorithm using the same underlying bipartite matching algorithm, 
because such opportunities for improving the quality of a certain matching usually do not all present 
themselves in a single designated time slot (for a regular switching algorithm to compute this matching).   
Intuitively, the larger the batch size $T$ is, the better the quality of a resulting matching is, since 
a larger $T$ provides a wider ``window of opportunities" for improving the quality 
of the matching as just explained.  

However, existing batch switching algorithms are not without shortcomings.  
They all suffer from at least one of the following two problems.
First, all existing batch switching algorithms except~\cite{Wang2018ParallelEdgeColoring} are 
serial algorithms and it is not known whether any of them can be parallelized.
As a result, they all have a time complexity of at least $O(N)$ per matching computation, 
since it takes $O(N)$ time just to ``print out" the computed result.  
This $O(N)$ time complexity is clearly too high for high-radix high-line-rate switches as just explained.  
Second, most existing switching algorithms require a large batch size $T$ to produce high-quality matchings 
that can lead to high throughputs, as will be elaborated in~\autoref{sec:related}. For example, 
it was reported in~\cite{Wang2018ParallelEdgeColoring} that the batch size had to be \num{3096} 
(for $N\!=\!300$ ports) for the algorithm to attain 96\% throughputs under some
traffic patterns.   A large batch size $T$ is certain to lead to poor delay performance: 
Regardless of the offered load condition, the average packet delay for any batch switching algorithm due to batching 
is at least $T/2$, since 
any packet belonging to the current batch has to wait till at least the beginning of the next batch to be switched.

\subsection{Our Contributions}

The first contribution of this work is a novel batch switching algorithm, called SB-QPS (Small-Batch QPS), 
that addresses both weaknesses of existing batch switching algorithms.  
First, it can attain a high throughout of over 85\%, under various traffic load patterns, 
using only a small batch size of $T = 16$ time slots.  This much smaller batch size translates into much better delay 
performances than those of existing batch switching algorithms, as will be shown in~\autoref{subsec:delay-results}.  
Second, SB-QPS is a fully distributed algorithm so that the matching computation load can be efficiently 
divided evenly across the $2N$ input and output ports.
As a result, its time complexity is the lowest possible:  $O(1)$ per matching computation per port.  

The design of the SB-QPS algorithm is extremely simple.  Only $T$ rounds of request-accept message exchanges 
by the input and the output ports are required for computing the $T$ matchings used (as the crossbar configurations) 
in a batch of $T$ time slots.  In each round, each input port $i$ sends a pairing request to an output port that is
sampled (by input port $i$) in a random queue-proportional fashion:  Each output port $j$ is sampled with a probability proportional to the
length of the corresponding VOQ. 
For this reason, we call this algorithm small-batch QPS (queue-proportional sampling).  
Since each QPS operation can be performed in $O(1)$ time using a simple data structure as shown in~\cite{Gong2017QPS}, 
the time complexity 
of SB-QPS is $O(1)$ per matching computation per port.  As will be explained in~\autoref{subsec:sb-qps-alg}, 
the way QPS is used in this work (SB-QPS) is very different than 
that in~\cite{Gong2017QPS}.  
For one thing, whereas in~\cite{Gong2017QPS} QPS is used as an auxiliary component to other switching algorithms such as iSLIP~\cite{McKeown99iSLIP} and SERENA~\cite{GiacconePrabhakarShah2003SERENA}, 
in this work, QPS serves the primary building block for SB-QPS.

Even though SB-QPS has a much smaller batching delay than other batch switching algorithms due to its much smaller $T$, 
the batching delay accounts for the bulk of the total packet delay under light to moderate traffic loads, when all other delays are
comparatively much smaller.  The second contribution of the work is to achieve the unthinkable:  a novel switching algorithm
called SW-QPS (SW for sliding window) that inherits and enhances all the good features of SB-QPS yet pays zero batching
delay.  More precisely, it has the same $O(1)$ 
time complexity as and achieves strictly better throughput and delay performances than SB-QPS.

SW-QPS does so by 
solving the switching problem under a novel framework
called {\it sliding-window switching}.  A sliding-window switching algorithm is different than a batch algorithm only in the following aspect.  
In a batch switching algorithm, a batch of $T$ matchings are produced every $T$ time slots.
In contrast, in a sliding-window switching algorithm, each window is still of size $T$ but 
a single matching is produced every time slot just like in a regular switching algorithm.   More precisely, at the beginning
of time slot $t$, the sliding window contains matchings-under-computation for the $T$ time slots $t$, $t+1$, $\cdots$, $t + T - 1$.   
The ``leading edge of the window", corresponding to the matching for the time slot $t$ (the ``senior class"), ``graduates" and is used as the crossbar configuration for the current time slot $t$.   
Then at the end of time slot $t$, a new and currently empty matching is added to the ``tail end of the window" as the ``freshman class".  
This matching 
will be computed in the next $T$ time slots and hopefully becomes a high-quality matching by the time $t + T$, when it ``graduates".  
SW-QPS completely removes the batching delay because ``it graduates a class every year" and furthermore always schedules an incoming packet to ``graduate"
at the earliest ``year" possible.

We consider SW-QPS to be the only research outcome of this work, since it strictly outperforms SB-QPS.  However, we describe both SB-QPS and SW-QPS
in detail for two reasons.  First, the incremental contributions of SB-QPS over existing batch switching algorithms and that of SW-QPS over SB-QPS are orthogonal 
to each other:  The former is to significantly reduce the batch size without sacrificing the throughput performance much and to reduce the time complexity to 
$O(1)$ via parallelization, whereas the latter is to retain the 
full benefits of batching without paying the batching delay.  Second, thanks to this orthogonality, explaining the 
differences between SB-QPS and existing batch switching algorithms and
that between SW-QPS and SB-QPS separately and incrementally makes the presentation much easier, as will become apparent in~\autoref{sec:sb-qps} and~\autoref{sec:sw-qps}.

The rest of this paper is organized as follows. In \autoref{sec:sys-module}, we state assumptions and the problem model used in this work. 
\autoref{sec:sb-qps} and \autoref{sec:sw-qps} detail the SB-QPS and SW-QPS algorithms respectively.  
In~\autoref{sec:related}, we survey the related works. 
Then, we evaluate the performances of SB-QPS and SW-QPS in~\autoref{sec:eva} and in \autoref{sec:con}, we conclude this paper. 

\section{Assumptions and Problem Model}\label{sec:sys-module}

In this work, we make the following two assumptions that are widely adopted in the 
literature ({\it e.g.,}~\cite{McKeown99iSLIP,Gong_QPSr_2019}). 
First, we assume that all incoming variable-length packets are first segmented into fixed-length packets, 
which are then reassembled before leaving the switch.  Hence, we consider the switching 
of only fixed-length packets in the sequel, and each such fixed-length packet takes exactly one time slot to transmit. 
Second, we assume that input ports, output ports and the crossbar operate at the same speed. 

An $N\times N$ input-queued crossbar can be modeled as a weighted bipartite graph, of which the
two disjoint vertex sets are the $N$ input ports and the $N$ output ports 
respectively.
We note that the edge set in this bipartite graph might change from a time slot to another. 
In this bipartite graph during a certain time slot $t$, there is an
edge between input port $i$ and output port $j$, if and only if
the $j^{th}$ VOQ at input port $i$,
the corresponding VOQ,
is nonempty (at $t$).  The weight of this edge is defined
as the length of ({\it i.e.,} the number of packets buffered at) this VOQ. 
A set of such edges constitutes a {\it valid crossbar schedule}, or a {\it matching},
if any two of them do not share a common vertex. 
\section{Small-Batch QPS} \label{sec:sb-qps}


\subsection{Batch Switching Algorithms}\label{subsec:back-batching}

\begin{figure}[!ht]
\centering
\includegraphics[width=0.60\columnwidth]{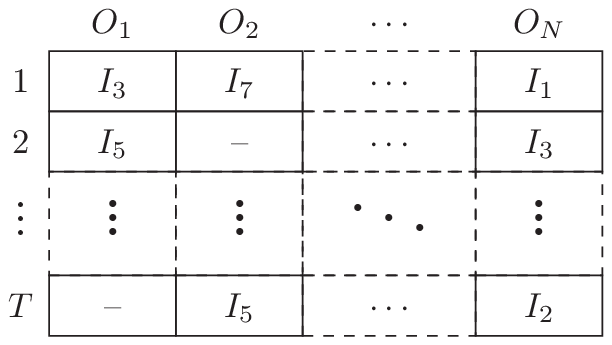}
\caption{A joint calendar. ``--" means unmatched.}\label{fig:sched-graph}
\end{figure}

Since Small-Batch QPS (SB-QPS) is a batch switching algorithm~\cite{Aggarwal2003EdgeColoring,Neely2007FrameBased,Wang2018ParallelEdgeColoring}, 
we first provide some background on batch switching. 
In a batch switching algorithm, the $T$ matchings for a batch of $T$ future time slots are batch-computed. 
These $T$ matchings form a joint calendar (schedule) of the $N$ output ports 
that can be encoded  
as a $T \times N$ table with $TN$ cells in it, as illustrated by an example shown in~\autoref{fig:sched-graph}.
Each column corresponds to the calendar of an output port and each row a time slot.  
The content of the cell at the intersection of the $t^{th}$ row and the $j^{th}$ column is the input port that $O_j$ is to pair with during 
the $t^{th}$ time slot in this batch.  Hence, each cell also corresponds to an edge (between the input and the output port pair)
and each row also corresponds to a matching (under computation for the corresponding time slot).   
In the example shown in~\autoref{fig:sched-graph},
output port $O_1$ is to pair with $I_3$ during the $1^{st}$ time slot (in this batch), $I_5$ 
during the $2^{nd}$ time slot, and is unmatched during the $T^{th}$ time slot.

At each input port, all packets that were in queue before a cutoff time (for the current batch), 
including those that belong to either the current batch, or previous batches but could not be served then,
are waiting to be inserted into the respective calendars ({\it i.e.,} columns of cells) of the corresponding output ports.  
The design objective of a batch switching algorithm is to pack as many such packets across the $N$ input ports as possible 
into the $TN$ cells in this joint calendar.   
After the computation of the current joint calendar is completed, the $T$ matchings in it will be used as the crossbar configurations for a batch of
$T$ future time slots.  In the meantime, the switch is switching packets according to the $T$ matchings specified 
in a past joint calendar that was computed earlier.

\subsection{The SB-QPS Algorithm}\label{subsec:sb-qps-alg}

In this section, we describe in detail SB-QPS, a batch switching algorithm that uses a small 
constant batch size $T$ that is independent of $N$. 
SB-QPS is a parallel iterative algorithm:  The input and output ports run 
$T$ QPS-like iterations (request-accept message exchanges) to collaboratively pack the joint calendar.     
The operation of each iteration is extremely simple:  Input ports request for cells in the joint calendar, 
and output ports accept or reject the requests.  More precisely, each iteration of SB-QPS, like that of QPS~\cite{Gong2017QPS}, 
consists of two phases: a proposing phase and an accepting phase. 

\noindent
{\bf Proposing Phase.} 
We adopt the same proposing strategy as in QPS~\cite{Gong2017QPS}. 
In this phase, each input port, unless it has no packet to transmit, proposes to {\it exactly one} output port 
that is decided by the QPS strategy. Here, we will only describe the operations at input port $1$; 
those at any other input port are identical. Like in~\cite{Gong2017QPS}, we denote by $m_1,m_2, \cdots, m_N$ 
the respective queue lengths of the $N$ VOQs at input port $1$, and by $m$ their sum ({\it i.e.}, $m \!\triangleq\!\sum_{k=1}^N m_k$). 
At first, input port 1 simply samples an output port $j$ with probability $m_j/m$, 
{\it i.e.,} proportional to $m_j$, the length of the corresponding VOQ; it then sends a proposal to output port $j$. 
The content of the proposal in SB-QPS is slightly different than that in QPS.  
In QPS, the proposal contains only the VOQ length information ({\it i.e.,} the value $m_j$),
whereas in SB-QPS, it contains also the following availability information (of input port $1$):  Out of the $T$ time slots in the batch,
what (time slots) are still available for input port $1$ to pair with an output port?
The time complexity of this QPS operation, 
carried out using the data structure proposed in~\cite{Gong2017QPS}, is $O(1)$ per input port.   

\noindent
{\bf Accepting Phase.} In SB-QPS, the accepting phase at an output port is quite different than that in QPS~\cite{Gong2017QPS}. 
Whereas the latter allows at most one proposal to be accepted at any output port (as QPS is a part of a regular switching algorithm that 
is concerned with only a single time slot at a time), 
the former allows an output port to accept multiple (up to $T$)  proposals (as each output port has up to $T$ cells in its calendar to be filled). 
Here, we describe the accepting phase at output port 1; that at any other output port is identical. 
The operations at output port 1 depend on the number of proposals it receives. If output port $1$ receives exactly one proposal from 
an input port (say input port $i$), it tries to accommodate this proposal using an accepting strategy we call {\it First Fit Accepting} (FFA).
The FFA strategy is to match in this case input port $i$ and output port $1$ at the earliest time slot (in the batch of $T$ time slots) 
during which both are still available (for pairing);  if they have ``schedule conflicts" over all $T$ time slots, this proposal is rejected. 
If output port $1$ receives proposals from multiple input ports, then it 
first sorts (with ties broken arbitrarily) these proposals in a 
descending order according to their corresponding VOQ lengths, and then tries to 
accept each of them using the FFA strategy.
 



In SB-QPS, opportunities -- in the form of proposals from input ports -- can arise, throughout the 
time window (up to $T$ time slots long) for computing the join calendar, to fill any of its $TN$ cells.  As explained earlier, 
this ``capturing every opportunity" to fill the joint calendar allows a batch switching algorithm to produce matchings of 
much higher qualities than a regular switching algorithm that is based on the same underlying bipartite matching algorithm can.   
Indeed, SB-QPS, the batch switching algorithm that is based on the QPS bipartite matching primitive, 
significantly outperforms QPS-1, the regular switching algorithm that is also based on QPS, as we will show in~\autoref{sec:eva}.


\noindent
{\bf Time Complexity.} 
The time complexity for the accepting phase at an output port is 
$O(1)$ on average, although in theory it can be as high as $O(N\log N)$ since an output port can receive up to $N$ 
proposals and have to sort them based on their corresponding VOQ lengths.  
Like in~\cite{Gong2017QPS}, this time complexity can be made $O(1)$ even in the worst case by letting the output port drop (``knock out") all proposals 
except the earliest few (say $3$) to arrive.
In this work, we indeed 
set this threshold to $3$ and find that it has a negligible effect on the quality of resulting matchings.    

We now explain how to carry out an FFA operation in $O(1)$ time.  
In SB-QPS, we encode the availability information of an input port $i$ as a $T$-bit-long bitmap $B_i[1..T]$, where $B_i[t]\!=\!1$ if input port $i$ is available (i.e., not 
already matched with an output port) at time slot $t$ and $B_i[t]\!=\!0$ otherwise.
The availability information of an output port $o$ is similarly encoded into a $T$-bit-long bitmap $B_o[1..T]$.  
When input port $i$ sends a proposal, which contains the availability information $B_i[1..T]$, to output port $o$, the corresponding FFA operation is for the output port $o$ to 
find the first bit in the bitmap $(B_i\&B_o)[1..T]$ that has value $1$, where ``$\&$" denotes bitwise-AND.  
Since the batch size $T$ in SB-QPS is a small constant 
(say $T$=16), both bitmaps can fit into a single CPU word and ``finding the first $1$" is an instruction on most modern CPUs.  

To summarize, the worst-case time complexity of SB-QPS is $O(T)$ per input or output port for the joint calendar consisting of $T$ matchings, 
since SB-QPS runs $T$ iterations and each iteration has $O(1)$ worst-case time complexity per input or output port.  Hence the worst-case 
time complexity for computing each matching is $O(1)$ per input or output port.




\noindent
{\bf Message Complexity.} The message complexity of each ``propose-accept'' iteration is $O(1)$ messages
per input or output port, because each input port sends at most one proposing message per iteration and each 
output port sends out at most 3 acceptance messages (where $3$ is the ``knockout" threshold explained above).  
Each proposing message is $T\!+\!\lceil\log_2W\rceil$ bits long ($T$ bits for encoding the availability 
information and $\lceil\log_2W\rceil$ bits for encoding the corresponding VOQ length), where $W$ is the longest possible VOQ length. 
Each acceptance message is  
$\lceil \log_2 T\rceil$ bits long (for encoding the time slot the pairing is to be made).  

\section{Sliding-Window QPS}\label{sec:sw-qps}

In this section, we present in detail the Sliding-Window QPS (SW-QPS) algorithm, the final and only research product of this work.  
Before we do so, we describe next the sliding-window framework that SW-QPS builds on.


\subsection{Sliding-Window Switching}\label{sec:sw-s}

\begin{figure}[!ht]
\centering
\includegraphics[width=0.95\columnwidth]{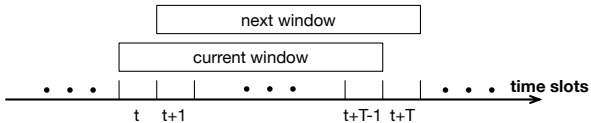}
\caption{Sliding-window switching.}\label{fig:sws}
\end{figure}

As mentioned earlier, the only difference between SW-QPS and SB-QPS is that SW-QPS changes the batch switching operation of SB-QPS to a 
sliding-window switching operation.  
Sliding-window switching combines regular switching with batch switching and gets the better of both worlds, as follows.
On one hand, during each time slot, under a sliding-window switching operation,  
there are $T$ matchings under computation, just like under a batch switching operation.
Each such matching has had or will have a window of $T$ time slots to find opportunities to have its quality improved 
by the underlying bipartite matching algorithm before it ``graduates''.
Hence, each such matching, when it ``graduates", can have a similar or even better quality than that computed by 
the batch switching algorithm that is based on the same 
underlying bipartite matching algorithm, as will be confirmed in~\autoref{sec:eva}. 

On the other hand, under a sliding-window switching operation, the ``windows of opportunities" of these $T$ matchings are staggered so that 
one matching (``class") is output (``graduated") every time slot.   This matching is to be used as the crossbar configuration for the current time slot. 
In this respect, it behaves like a regular switching algorithm and hence 
completely eliminates the batching delay of the batch switching.    
More specifically, at the beginning of time slot $t$, the most senior matching (``class") 
in the window was added (``enrolled") to the window at the end of time slot $t - T - 1$ and is to ``graduate" at the beginning of time slot $t$, so its ``window of opportunity"
(to have its quality improved) is $[t - T, t-1]$.  The ``window of opportunity" for the second most senior matching is $[t - T + 1, t]$ and so on.  
At the end of time slot $t$, a ``freshman class'' (an empty matching) is ``enrolled" and scheduled to ``graduate" at time slot $t + T + 1$ in the future.  


\autoref{fig:sws} shows how the sliding window evolves from time slot $t$ to time slot $t+1$.  In~\autoref{fig:sws}, each interval along the timeline corresponds to 
a ``class".   
As shown in~\autoref{fig:sws}, at the beginning of time slot $t$, the current window contains 
``classes of years" (matchings-under-computation to be used as crossbar schedules for time slots) 
$t$, $t+1$, $\cdots$, and $t + T - 1$.  Then, at the beginning of time slot $t+1$, the current window slides right by $1$ (time slot), and the
new window contains ``classes of years" $t+1$, $t+2$, $\cdots$, and $t+T$,
because the ``class of year $t$" just graduated and the ``class of year $t+T$" was just ``enrolled".   

In theory, almost any batch switching algorithm can be converted into a sliding-window switching algorithm by making the ``windows of opportunity" for 
the batch of $T$ matchings-under-computation staggered instead of aligned.   This conversion would in general improve switching performance by eliminating the batching delay. 
Hence, this sliding-window switching framework is itself a separate
contribution of this work.

\subsection{The SW-QPS Algorithm} \label{subsec:sw-qps-alg} 



SW-QPS is exactly such a conversion of the batch switching algorithm SB-QPS into a sliding-window switching algorithm.  
SW-QPS is also a parallel iterative algorithm whose each iteration is identical to that of SB-QPS.  Hence SW-QPS 
has the same $O(1)$ time and $O(1)$ message complexities (per port per matching computation)
as SB-QPS.  
The only major difference is that, SW-QPS ``graduates'' a matching every time slot whereas SB-QPS ``batch-graduates" $T$ matchings every $T$ time slots. 
This ``graduating a class each year" allows SW-QPS to completely eliminate the batching delay.
As explained earlier, in SW-QPS, at the beginning of time slot $t$, the joint calendar consists of the $T$ matchings-under-computation that are to ``graduate" in 
``years" (time slots) $t$, $t+1$, $\cdots$, $t + T - 1$ respectively.  Hence at time slot $t$, the $T$-bit-long availability bitmap of an input port $i$ indicates the availabilities of 
$i$ during $[t, t + T -1]$.   

Note that SW-QPS inherits the FFA (First Fit Accepting) strategy of SB-QPS that is to arrange for an input-output pairing -- 
and hence the switching of a packet
between the pair -- at the earliest mutually available time slot.  In other words, an incoming packet is always ``advanced to the most senior class that it can fit in
schedule-wise" so that it can ``graduate'' at the earliest ``year'' possible. 
This greedy strategy further reduces the queueing delay of a packet, as will be shown 
in~\autoref{sec:eva}.

\section{Related Work}\label{sec:related}

In this section, we provide a brief survey of prior studies that are directly related to ours. 

\noindent
{\bf Regular Switching Algorithms.} 
Using MWM (Maximum Weighted Matching) as crossbar schedules
is known to result in 100\% switch throughput and
near-optimal 
queueing delays under various traffic patterns~\cite{McKeownMekkittikulAnantharamEtAl1999}, but
each MWM takes $O(N^{2.5}\log W)$ time to compute using the state-of-the-art algorithm~\cite{Duan2012BpMWMScaling}, where
$W$ is the maximum possible length of a VOQ. Motivated by this, 
various parallel exact or approximate 
MWM algorithms ({\it e.g.,}~\cite{Fayyazi04ParallelMWM, BayatiPrabhakarShahEtAl2007Iterative}) have been proposed to reduce its time complexity. 
However, the time complexities of all these algorithms above are still too high to be used in high-line-rate high-radix switches. 

The family of parallel iterative algorithms~\cite{McKeown99iSLIP,Hu_HighestRankFirst_2018,Hu2016IterSched,Gong_QPSr_2019} generally has a low time complexity per port. However, 
their throughput and delay performances are generally much worse than those of MWM. 
We note that QPS-r~\cite{Gong_QPSr_2019}, the state-of-the-art algorithm in this family, also
builds on QPS~\cite{Gong2017QPS}.  It simply runs $r$ (a small constant) iterations of QPS to arrive at a final matching. 
We will compare our SB-QPS and SW-QPS with it in~\autoref{sec:eva}.

\noindent
{\bf Batch Switching Algorithms.}
Most of the existing batch switching algorithms~\cite{Aggarwal2003EdgeColoring, Neely2007FrameBased, Wang2018ParallelEdgeColoring} 
model the process of packing the joint calendar as an edge-coloring problem, but until now, 
most practical solutions to the latter problem are centralized and have high complexity. 
For example, the Fair-Frame algorithm~\cite{Neely2007FrameBased} based on the Birkhoff von Neumann Decomposition (BvND) 
has a time complexity of $O(N^{1.5}\log N)$ per matching computation. 

A recent work, based on parallel edge coloring, has been proposed in~\cite{Wang2018ParallelEdgeColoring}. 
It pushes the per-port time complexity (per matching computation) down to $O(\log N)$.
It requires a bath size of only $O(\log N)$, but as mentioned in~\autoref{sec:intro}, the constant factor hidden in 
the big-O is very large.

\section{Performance Evaluation}\label{sec:eva}

In this section, we evaluate, through simulations, the throughput and delay performances of SB-QPS and SW-QPS 
under various load conditions and traffic patterns.  Our algorithms are compared against 
iSLIP~\cite{McKeown99iSLIP}, which runs $\log_2 N$ request-grant-accept iterations and is hence much more
expensive computationally.
Our algorithms are also compared against QPS-1 
(QPS-r with $r$=1 iteration)~\cite{Gong_QPSr_2019}.  This is a fair comparison because QPS-1, like our 
algorithms, runs only a single iteration to compute a matching.  
The MWM algorithm, which delivers near-optimal delay performance~\cite{ShahWischik2006MWM0}, 
is also compared against as a benchmark. 



\subsection{Simulation Setup}
\label{subsec: sim-setting}

In our simulations, we 
fix the number of input and output ports $N$ to $64$;  we however will investigate in \autoref{app-sec:delay-vs-port}
how the mean delay performances of these algorithms scale with respect to $N$.
To accurately measure throughput and delay, we assume that each VOQ has an infinite buffer size, so no packet is dropped at any input port. Each simulation run follows the stopping rule in \cite{Flegal2010MCStopRule, Glynn1992MCStopRule}:   The number of time slots simulated is at least $500N^2$ and guarantees the difference between the estimated and the actual average delays to be within $0.01$ time slots with at least $0.98$ probability.

We assume in our simulations that each traffic arrival matrix $A(t)$ is {\it i.i.d.} Bernoulli with its traffic rate matrix  equal to the product of the offered load and a traffic pattern
matrix (defined next). Similar Bernoulli arrivals were studied in  ~\cite{GiacconePrabhakarShah2003SERENA,McKeown99iSLIP,Gong2017QPS}. 
Later, in~\autoref{app-sec: bursty-arrivals-bench}, we will look at burst traffic arrivals. 
Note that only synthetic traffic (instead of that derived from packet traces) is used in our simulations
because, to the best of our knowledge, there is no meaningful way to combine
packet traces into switch-wide traffic workloads. 
The following four standard types of normalized (with each row or column sum equal to $1$) traffic patterns are used:
(I) \emph{Uniform}: packets arriving at any input port go
to each output port with probability $\frac{1}{N}$.
(II) \emph{Quasi-diagonal}: packets arriving at input port $i$ go to
output port $j \!=\! i$ with probability $\frac{1}{2}$ and go to any other output port
with probability $\frac{1}{2(N-1)}$.
(III) \emph{Log-diagonal}: packets arriving at input port $i$ go
to output port $j = i$ with probability $\frac{2^{(N-1)}}{2^N - 1}$ and
go to any other output port $j$ with probability equal $\frac{1}{2}$ of the
probability of output port $j - 1$ (note: output port $0$ equals output port $N$).
(IV) \emph{Diagonal}: packets arriving at input port $i$ go to
output port $j \!=\! i$ with probability $\frac{2}{3}$, or go to output port
$(i\, \text{mod} \, N) + 1$ with probability $\frac{1}{3}$.
These traffic patterns are listed in order of how skewed the volumes of traffic arrivals to different output
ports are: from uniform being the least skewed, to diagonal being the most skewed.

When implementing SB-QPS and SW-QPS, we have to first decide on the value of batch (for SB-QPS) or window (for SW-QPS) size $T$. 
As explained earlier in~\autoref{sec:intro}, 
for SB-QPS, a larger batch size $T$ generally results in matchings of higher qualities and hence leads to better throughput performances.
However, a larger $T$ results in longer batching delays and hence can lead to worse overall delay performances for SB-QPS.     
In addition, since the availability information in a proposal message is $T$ bits long, a larger $T$ leads to a higher communication complexity for SB-QPS.  
Through simulations (results not shown here in the interest of space), 
we have found that $T = 16$ strikes a nice performance-cost 
tradeoff:   The batching delay is reasonably low and the proposal message size is
small when $T = 16$, yet the throughput gains when increasing $T$ beyond 16 (say to 32) are marginal for SB-QPS.   
Hence we set $T = 16$ for SB-QPS.  SB-QPS clearly deserves its  
name (small-batch) since this tiny batch size of 16 is much smaller than 
that of any other batch switching algorithm.  

Since SW-QPS completely eliminates the batching delay, the only cost of increasing $T$ for SW-QPS is the larger proposal message size.  
Nonetheless, we have found that $T = 16$ is a nice performance-cost tradeoff point, and hence is adopted, also for SW-QPS.  
For SW-QPS, $T$ does not have to grow with $N$ (to deliver similar throughput and delay performances),
as we will show in~\autoref{app-sec:delay-vs-port} that the delay performance of SW-QPS (with $T = 16$) 
does not degrade when $N$ grows larger.

\begin{figure*}
    \centering
    \includegraphics[width=\fourfull]{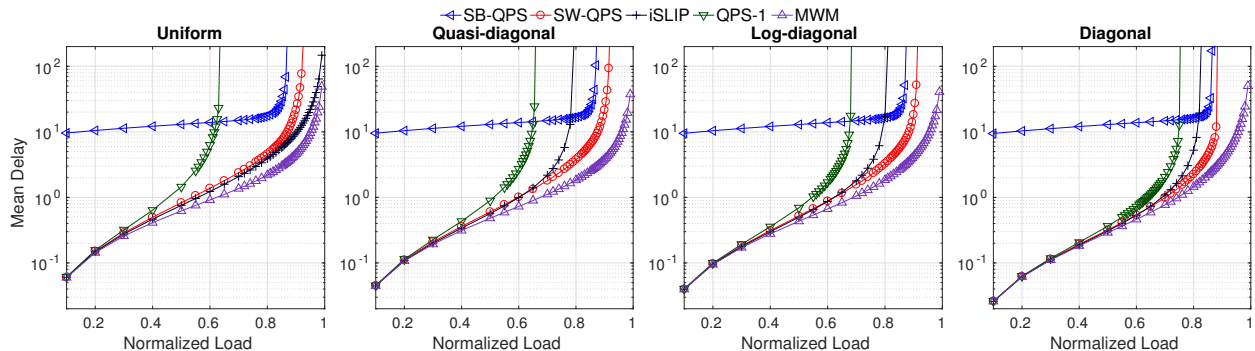}
    \caption{Mean delays of SB-QPS, SW-QPS, iSLIP, QPS-1, and MWM under the $4$ traffic patterns.}\label{fig: delay-load-bench}
\end{figure*}

\subsection{Throughput Performance Results}\label{subsec:throughput-results}

\begin{table}[ht!]
    \centering
    \caption{Maximum achievable throughput.}\label{tab:thru-comp}
    \begin{tabular}{@{}|l|c|c|c|c|@{}}
        \hline
        \textbf{Traffic} &  Uniform  & Quasi-diag  &  Log-diag & Diag  \\
        \hline
        \textbf{SB-QPS}& 86.88\% & 87.10\% & 87.31\% & 86.47\% \\
        \hline
        \textbf{SW-QPS} & 92.56\% & 91.71\% & 91.40\% & 87.74\% \\
        \hline 
        \textbf{iSLIP} & 99.56\% & 80.43\% & 83.16\% & 82.96\% \\
        \hline
        \textbf{QPS-1} & 63.54\% & 66.60\% &  68.78\% &75.16\%\\
        \hline
    \end{tabular}%
\end{table}

\autoref{tab:thru-comp} presents the maximum achievable throughput of SB-QPS, SW-QPS, iSLIP, and QPS-1, under 
the aforementioned four standard traffic patterns and an offered load close to $1$ (more precisely, 0.9999). 
We do not include the throughout of MWM in~\autoref{tab:thru-comp}, because it can provably attain $100\%$ throughput. 
We make three observations from \autoref{tab:thru-comp}.  First, SW-QPS significantly 
improves the throughput performance of QPS-1, increasing it by an additive term of 
0.2902, 0.2511, 0.2262, and 0.1258 for the uniform, quasi-diagonal, log-diagonal, and diagonal traffic patterns respectively. 
Second, the throughput of SW-QPS is consistently higher than that of SB-QPS under the four traffic patterns. 
Third, under all traffic patterns except uniform, SW-QPS significantly outperforms iSLIP, which is much more expensive computationally as it runs $\log_2 N$ iterations for each matching computation. 

\subsection{Delay Performance Results}\label{subsec:delay-results}

\autoref{fig: delay-load-bench} shows the mean delays of SB-QPS, SW-QPS, iSLIP, QPS-1, and MWM 
under the aforementioned four traffic patterns.  As we have shown in~\autoref{subsec:throughput-results}, 
SB-QPS, SW-QPS, iSLIP and QPS-1 generally cannot attain 100\% throughput, so we only measure their delay performances for  
the offered loads under which they are stable; in all figures in the sequel, each ``missing point'' on a plot indicates 
that the corresponding algorithm is not stable under the corresponding traffic pattern and offered load. 
\autoref{fig: delay-load-bench} shows that, when the offered load is not very high (say $<0.6$), 
SB-QPS has a much higher mean overall delay than others thanks to its batching delay that is still relatively 
quite high (despite a small batch size of $T = 16$);  in comparison, SW-QPS completely eliminates this batching delay. 
\autoref{fig: delay-load-bench} also shows that SW-QPS outperforms QPS-1 everywhere and outperforms 
iSLIP under all traffic patterns except uniform.  
Since as shown in~\autoref{tab:thru-comp} and~\autoref{fig: delay-load-bench}, the throughput and the delay performances 
of SW-QPS are strictly better than those of SB-QPS, we will show the performance results of only SW-QPS in~\autoref{app-sec:more-results}, 
in which we present more evaluation results.

\section{Conclusion}\label{sec:con}

In this work, we first propose a batch switching algorithm called SB-QPS that significantly reduces the batch size without 
sacrificing the throughput performance much, and achieves 
a time complexity of $O(1)$ 
per matching computation per port via parallelization. 
We then propose a regular switching algorithm called SW-QPS that improves on SB-QPS using a novel sliding-window switching framework. 
SW-QPS inherits and enhances all benefits of SB-QPS and reduces the batching delay to zero. 
We show, through 
simulations, that the throughput and delay performances of SW-QPS are much better than those of QPS-1, the state-of-the-art regular switching algorithm 
based on the same underlying bipartite matching algorithm.

%

\smallskip
\noindent
{\bf Acknowledgments.}
This material is based upon work supported by the National Science Foundation under Grant No. CNS-1909048 and CNS-2007006.



	%
\bibliographystyle{abbrv}
\bibliography{bibs/ref}

\appendix

\begin{figure*}
 	\begin{minipage}{\textwidth}
	    \centering
    \includegraphics[width=\fourfull]{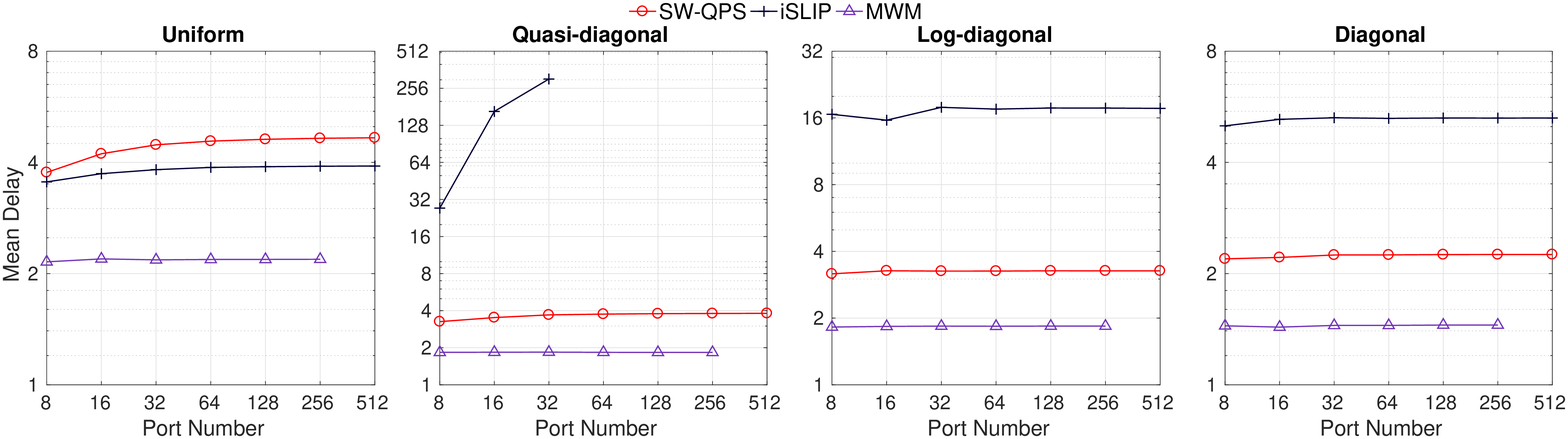}
      \caption{Mean delays v.s. number of (input/output) ports under
  i.i.d.~Bernoulli traffic arrivals (offered load: $0.8$).}\label{fig: delay-port}
	\end{minipage}
	\begin{minipage}{\textwidth}
	\centering
	\includegraphics[width=\fourfull]{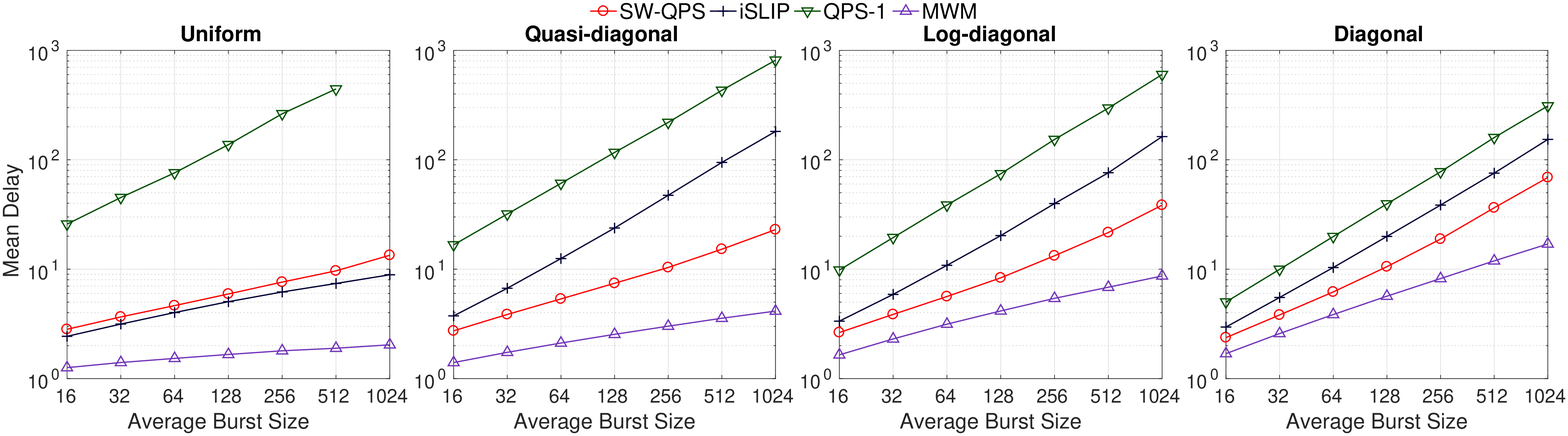}
	\caption{Mean delays of SW-QPS, iSLIP, QPS-1, and MWM under bursty arrivals (offered load: $0.6$).}
	\label{fig: delay-burst60}
	\end{minipage}
\end{figure*}

\section{More Evaluation Results}\label{app-sec:more-results}

\subsection{How Mean Delay Scales with N}\label{app-sec:delay-vs-port}
In this section, we investigate how the mean delays of SW-QPS, iSLIP, QPS-1, and MWM scale with the number of input/output 
ports $N$ under (non-bursty) i.i.d. Bernoulli traffic. 
We have simulated seven different $N$ values: $N\!=\!8, 16, 32, 64, 128, 256, 512$. 
We have simulated various offered loads, but here we only present the results 
under an offered load of $0.8$; other offered loads (say $0.6$) lead to similar conclusions.
\autoref{fig: delay-port} shows the simulation results, under the $4$ different traffic patterns 
under an offered load of $0.8$. The results of QPS-1 are not shown, because it is not stable when the offered load is $0.8$ under all four traffic patterns. 
Some points for iSLIP are missing because iSLIP is not stable when $N\!\ge\!64$ under the quasi-diagonal traffic pattern. 
The missing points for MWM are due to the fact we are not able to obtain its mean delays when $N$=512 in a 
reasonable amount of time (thanks to the high time complexity of the MWM algorithm). 
\autoref{fig: delay-port} shows that mean delays of 
SW-QPS, like those of MWM,
are almost independent of $N$.  We have also found through simulations that the maximum achievable throughputs of SW-QPS are also almost independent of $N$.

\subsection{Bursty Arrivals}\label{app-sec: bursty-arrivals-bench}

In real networks, packet arrivals are likely to be bursty.
In this section, we evaluate the performances of SW-QPS, iSLIP, QPS-1, and MWM under bursty traffic, 
generated by a two-state ON-OFF arrival process. 
The durations of each ON (burst) stage and OFF (no burst) stage are
geometrically distributed: the probabilities that the ON and OFF states last for $t \ge 0$ time slots are given by
$P_{ON}(t) = p(1-p)^t \text{ and } P_{OFF}(t) = q(1-q)^t$,
with the parameters $p, q \in (0,1)$ respectively. As such, the average duration of the
ON and OFF states are $(1-p)/p$ and $(1-q)/q$ time slots
respectively.

In an OFF state, an incoming packet's destination (\ie output
port) is generated according to the corresponding traffic pattern. In an
ON state, all incoming packet arrivals to an input port would be destined
to the same output port, thus simulating a burst of packet
arrivals. By controlling $p$, we can control the desired
average burst size while by adjusting $q$,
we can control the load of the traffic.

We evaluate the mean delay performances of these four algorithms, with the average burst size ranging 
from $16$ to \num{1024} packets, under a moderate offered load of $0.6$ and a heavy offered load of $0.8$, respectively. 
The simulation results for the former are shown in~\autoref{fig: delay-burst60}; 
those for the later are omitted, since they are similar except that iSLIP is not stable for some, and QPS-1 is not stable for all, average burst sizes under the offered load of $0.8$. 
One point for QPS-1 is missing in the leftmost sub-figure in~\autoref{fig: delay-burst60}, because QPS-1 is not stable when the average burst size becomes \num{1024} under the uniform traffic pattern and an offered load of $0.6$. 
\autoref{fig: delay-burst60} clearly shows that SW-QPS outperforms iSLIP (under all traffic patterns except uniform), 
and QPS-1 (under all traffic patterns) by an increasingly wider margin in both 
absolute and relative terms as the average burst size becomes larger. 

\end{document}